\documentclass[twocolumn,aps,pra,showpacs,amsfonts,amssymb,amsmath,a4paper,flequ]{revtex4-1}
\usepackage{mathrsfs}
\usepackage{amsmath}
\usepackage{subfigure}
\usepackage{float}
\usepackage[title,titletoc,header]{appendix}
\usepackage{graphicx}

\begin{document}
 \draft
 \title{ Zitterbewegung effect in spin-orbit coupled spin-1 ultracold atoms}
 \author{Yi-Cai Zhang$^{1}$, Song-Wei Song$^{1}$, Chao-Fei Liu$^{1,2}$ and Wu-Ming Liu$^{1}$}
 \address{$^{1}$Beijing National Laboratory for Condensed Matter
 Physics, Institute of Physics, Chinese Academy of Sciences, Beijing
 100190, China\nonumber\\
$^{2}$School of Science, Jiangxi University of Science and Technology, Ganzhou 341000, China}

\date{\today}

\begin{abstract}
The Zitterbewegung effect in spin-orbit coupled spin-1 cold atoms is investigated in the presence of the Zeeman
field and a harmonic trap. It is shown that the Zeeman field and the harmonic trap have significant effect on the Zitterbewegung oscillatory behaviors. The external Zeeman field could suppress or enhance the Zitterbewegung amplitude and change the frequencies of oscillation. A much slowly damping Zitterbewegung oscillation can be
achieved by adjusting both the linear and quadratic Zeeman field. Multi-frequency Zitterbewegung oscillation can be induced by the applied Zeeman field. In the presence of the harmonic trap, the subpackets corresponding to different eigenenergies would always
keep coherent, resulting in the persistent Zitterbewegung oscillations. The Zitterbewegung oscillation would display very complicated and irregular oscillation characteristics due to the coexistence of different frequencies of the Zitterbewegung oscillation. Numerical results show that, the Zitterbewegung effect is robust even in the presence of interaction between atoms.

\end{abstract}

\maketitle

\section{Introduction}
The Zitterbewegung (ZB) effect, which is characterized by high frequency oscillations (trembling motion) for Dirac electrons was firstly predicted by Schr\"{o}dinger \cite{Schrodinger}. The Zitterbewegung effect is purely a relativistic phenomenon and originates from the interference between the positive and negative energy states of the electron. The experimental observation of the electron Zitterbewegung has not been realized due to its  high frequency (the order of $\hbar \omega=2m_e c^2\sim1MeV$) and small oscillatory amplitude (the order of Compton wavelength of electron $\hbar/m_e c\sim10^{-12}m$). However, it is shown that there exist Zitterbewegung-like effects in graphene \cite{Rusin2,Schliemann1}, superconductors \cite{Jozsef}, photonic crystal \cite{Zhang2}, single trapped ion \cite{Lamata,Gerritsma} and semiconductor quantum wells \cite{Schliemann,Jiang}. Hence, the Zitterbewegung-like effect has attracted great attentions recently, both theoretically and experimentally in various physics fields \cite{Shen,Winkler,Lee,Wang}.



Most of these studies of the Zitterbewegung effect are based on the Dirac-like equation with spin-orbit coupling interaction.  The spin-orbit coupling interaction plays an essential role in  a lot of interesting  phenomena, such as quantum spin Hall effects, topological insulator, exotic superconductivity or superfluidity, etc. In recent years, the spin-orbit coupling interaction in cold atoms have attracted great attentions both experimentally and theoretically.  The spin-orbit coupling in two component atoms have been created experimentally with Raman laser beams \cite{Lin,Lin1,Wang2,Cheuk}. The proposal to realize spin-orbit coupling for three component atoms has been put forward in Ref. \cite{Juzeliunas}. Using the so-called tetrapod setup scheme and two pair of counterpropagating laser beams, the spin-orbit coupling in spin-1 atoms  could be realized in alkali-metal atoms. The resulting gauge potential is proportional to the projection of angular momentum operator of spin-1 atoms along $xy$ plan. The ground states of the spin-orbit coupled BEC have been extensively studied theoretically. For example, in the case of strong coupling and the weak harmonic trap, the spin-orbit coupling would result in nontrivial ground state in BEC, such as plane wave phase or standing wave phase, which depends on interaction between atoms \cite{Zhai,Ho,Yip}. For the strong coupling and the strong harmonic trap, the half-quantum vortex phase or vortex lattice phase develops \cite{Hu,Santos,Xu,Zhou,Guo}.

The length and energy scales can be well controlled in cold atoms. Thus, it is possible to observe the Zitterbewegung-like oscillation in cold atom experiments. Recently, there were proposals to simulate the Zitterbewegung effect by using spin-orbit coupled ultracold atoms \cite{Vaishnav,Zhang,Zhang1,Merkl}, wherein most authors focus on  the Zitterbewegung effect of two component atoms with spin-orbit coupling in free space. Thus, the quasi-momenta is a good quantum number and the Hamiltonian can be diagonalized within the momentum space. By using the Gaussian packet as the initial state, it is shown that the amplitude of the Zitterbewegung oscillation deceases with time and the Zitterbewegung phenomenon has a transient characteristic. It would be of interest to find ways to stabilize and manipulate the Zitterbewegung oscillation. Zitterbewegung-like phenomena are fairly common  characteristics for multi-level systems \cite{David}. However, the multi-frequency Zitterbewegung oscillation in ultracold atomic system has not been investigated to the best of our knowledge. The effects which is induced by the Zeeman field and the trap on the Zitterbewegung oscillations need to be clarified.

In the present paper, we investigate the characteristics of the multi-frequency oscillation induced by Zeeman field and the persistent oscillations in the presence of the external harmonic trap in the spin-1 ultra-cold atoms. In Sec. \textbf{II},  we introduce the general Hamiltonian of spin-1 atoms with spin-orbit coupling, and then focus on the Zitterbewegung oscillatory characteristics in the Zeeman fields. In Sec.
\textbf{III}, the Zitterbewegung oscillatory characteristics in harmonic trap are investigated.  In Sec.
\textbf{IV}, the effect of the interaction between atoms on the  Zitterbewegung oscillation is considered. A summary is presented in Sec. \textbf{V}.

\begin {figure}[H]
\centering
\includegraphics[ scale=0.42]{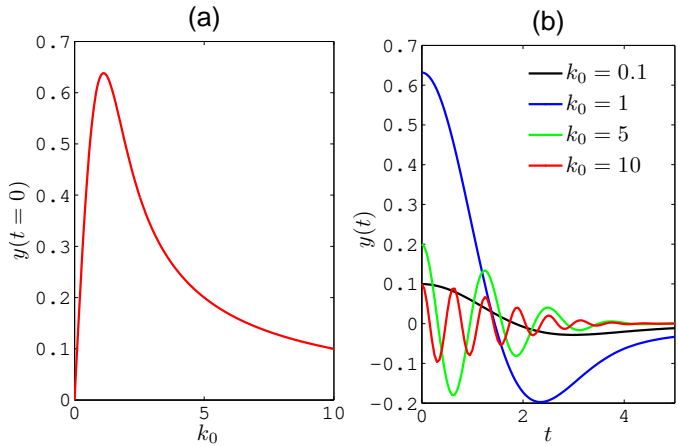}\\[0pt]
\caption{(color online). (a) The amplitude of the Zitterbewegung oscillation as a function of the momentum $k_0$ under zero Zeeman field. (b) The Zitterbewegung oscillations under zero Zeeman field for various momentum $k_0$.    }
\label{sw2d}
\end{figure}

\section{Zitterbewegung effect in Zeeman fields }
The two dimensional Hamiltonian of the spin-orbit coupled spin-1 atoms in the presence of the external Zeeman field and harmonic trap is \cite{Zhai}
\begin{align}
& H=T+V_{trap}+V_{SO}+V_{Z}+H_{int}, \notag \\
& T=\int d\vec{r}\Psi^{+}\frac{p_{x}^{2}+p_{y}^{2}}{2m}\Psi, \notag \\
& V_{trap}=\int d\vec{r}\Psi^{+}\frac{m\omega^2(x^{2}+y^{2})}{2}\Psi,  \notag \\
& V_{SO}=\int d\vec{r}\Psi^{+}[\gamma p_{x}F_{x}+\gamma p_{y}F_{y}]\Psi, \notag\\
& V_{Z}=\int d\vec{r}\Psi^{+} [pF_{z}+qF_{z}^{2}]\Psi,\notag\\
&H_{int}=\int d\vec{r}[\frac{c_0}{2}\hat{n}^{2}+\frac{c_2}{2}\textbf{F}^{2}],
\end{align}
where T, $V_{trap}$, $V_{SO}$,  $V_{Z}$, $H_{int}$ are the kinetic energy, the harmonic potential, the spin-orbit coupling interaction, the effective Zeeman shift and the interaction between atoms, respectively. $\Psi=(\Psi_1, \Psi_0, \Psi_{-1})^{T}$, $\hat{n}=\hat{n}_1+\hat{n}_0+\hat{n}_{-1}$, $\textbf{F}=\Psi^{+}_\alpha \vec{F}_{\alpha \beta}\Psi_\beta$, $c_0=4\pi\hbar^2(a_0+2a_2)/3m$ and $c_2=4\pi\hbar^2(a_2-a_0)/3$ are the field operator, the number density operator, the angular momentum operator, the spin-independent interaction parameter and the spin-dependent interaction parameter, respectively. $a_0$ and $a_2$ are the s-wave scattering lengths corresponding to the total spin of the two colliding bosons $0$ and $2$, respectively.
 The spin-orbit coupling strength and the effective Zeeman parameters can be tuned by the wave length of Raman lasers and the detuning from Raman resonance, respectively \cite{Lin1,Cheuk}.

\begin {figure}[H]
\centering
\includegraphics[ scale=0.55 ]{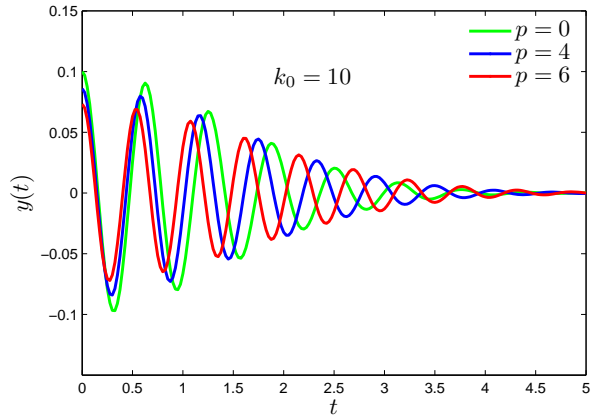}\\[0pt]
\caption{(color online). Zitterbewegung oscillation in spin-1 atoms under the linear Zeeman field. The Zitterbewegung oscillation under zero Zeeman field is plotted in green. The blue, red lines are Zitterbewegung oscillation under a linear Zeeman field p=4 and p=6, respectively. }
\label{sw2d}
\end{figure}

The Pauli spin matrices for spin spin-$1$ atoms are given by
\begin{align}
F_{x}=\left(                 
 \begin{array}{ccc}   
   0    &    \frac{1}{\sqrt{2}}    &  0 \\  
   \frac{1}{\sqrt{2}}   &    0    &  \frac{1}{\sqrt{2}}      \\  
   0     &   \frac{1}{\sqrt{2}}   &   0
\end{array}\right) ,
F_{y}=\left(                 
 \begin{array}{ccc}   
   0    &    \frac{-i}{\sqrt{2}}    &  0 \\  
   \frac{i}{\sqrt{2}}   &    0    &  \frac{-i}{\sqrt{2}}      \\  
   0     &   \frac{i}{\sqrt{2}}   &   0   \notag
\end{array}\right),
\end{align}

\begin{align}
F_{z}=\left(                 
 \begin{array}{ccc}   
   1    &    0    &  0 \\  
   0 &    0    &  0    \\  
   0     &  0   &   -1 \notag
\end{array}\right).
\end{align}

Without the external harmonic trap and interaction, the Hamiltonian can be diagonalized in the momentum space as
\begin{equation}
H'=\left(                 
 \begin{array}{ccc}   
   \omega_{1}    &    0    &  0 \\  
   0    &    \omega_{2}    &  0      \\  
   0     &   0   &  \omega_{3}     \\
\end{array}\right),
\end{equation}
where $\omega_{1}=({p_{x}^{2}+p_{y}^{2}})/2m+n\texttt{cos}{u}-b/3$, $\omega_{2}=({p_{x}^{2}+p_{y}^{2}})/{2m}+n\texttt{cos}(u+{4\pi}/{3})-{b}/{3}$,
$\omega_{3}=({p_{x}^{2}+p_{y}^{2}})/{2m}+n\texttt{cos}(u+{2\pi}/{3})-{b}/{3}$ are roots of eigen-equation $Det(\omega I-H)=0$. The corresponding eigenvectors are
\begin{equation}
|\alpha_{i}\rangle=\frac{1}{n_i}\left(                 
 \begin{array}{ccc}   
   -(p_{x}^{2}+p_{y}^{2})+2\omega_{i}(p-q+\omega_{i})    \\  
   \sqrt{2}(p-q+\omega_{i})(p_{x}+ip_{y})         \\  
   (p_{x}+ip_{y})^{2}           \\
\end{array}\right),
\end{equation}
where $n_i$ is normalization coefficient, $n=\sqrt{{-4p_1}/{3}}$, $u={\texttt{arccos}({-q_1({-p_1}/{3})^{{-3}/{2}}}/{2})}/{3}$, $p_1=c-{b^{2}}/{3}$, $q_1=d-{bc}/{3}+{2b^{3}}/{27}$, $b=-2q$, $c=-(p_{x}^{2}+p_{y}^{2})+q^{2}-p^{2}$, $d=(p_{x}^{2}+p_{y}^{2})q$, respectively. In general,
they are the functions of the momentum ($p_{x}$, $p_{y}$) and Zeeman parameters ($p$, $q$).

\begin{figure}[tbp]
\begin{center}
\includegraphics[ scale=0.55 ]{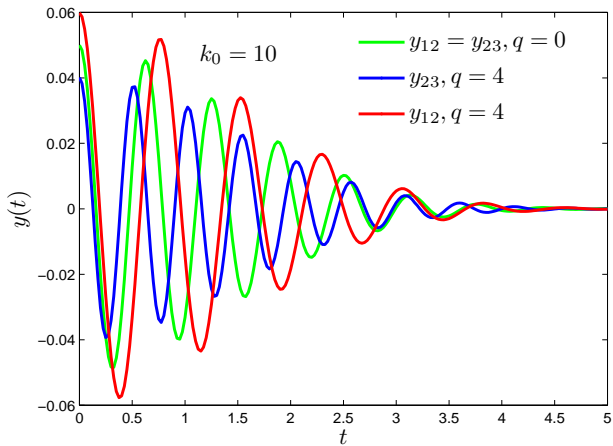}\\[0pt]
\end{center}
\caption{(color online). Zitterbewegung oscillation under the quadratic Zeeman field. The blue and red lines are Zitterbewegung oscillations corresponding to $\omega_{23}$ and $\omega_{12}$ for a quadratic Zeeman field parameter q=4. For the sake of comparison, the green line corresponding to zero Zeeman field is also plotted.}
\label{sw2d}
\end{figure}

The position operator in the Heisenberg picture is  \cite{David}
\begin{align}
& \vec{r}(t)=\vec{r}(0)+\sum_{k}Z_{k;k}+t\sum_{k}\vec{V}_{k}Q_{k}+\sum_{k,l\neq m}e^{i\frac{\omega_{kl}t}{\hbar}}\vec{Z}_{k;l},
\end{align}
where $\vec{V}_{k}=\frac{\partial \omega_{k}}{\partial \vec{p}}$, $Q_{k}=|\alpha_{k}\rangle\langle\alpha_{k}|$ are the group velocity and the projection operator, respectively. $k$, $l$ denote the energy branch $\omega_1$, $\omega_2$, $\omega_3$ and $\omega_{kl}=\omega_{k}-\omega_{l}$ is the eigen-energy difference. $Z_{k;l}=iQ_{k}\frac{\partial Q_{l}}{\partial \vec{p}}$ is the so-called $Zitterbewegung$ $amplitudes$ (see Ref. \cite{David}).  The first and second terms are constants in Eq. (4). The third term  is the uniform motion and the fourth one corresponds to the Zitterbewegung oscillation. As shown in the fourth term, the position operator usually undergoes a multi-frequency oscillation. We will calculate the average value of position by using an initial wave function of Gaussian density distribution:

\begin{align}
|g\rangle=\frac{1}{\sqrt{\pi\delta^{2}}}e^{-\frac{x^2+y^2}{2\delta^2}}e^{\frac{ik_{0}x}{\hbar}}\left(                 
  \begin{array}{ccc}   
    1\\  
    0\\  
    0\\  
  \end{array}
\right).                
\end{align}
 The wave function expressed in momentum space is
\begin{align}
|g\rangle=\sqrt{\frac{\delta^{2}}{\pi}}e^{-\frac{\delta^2}{2\hbar^{2}}[(p_{x}-k_{0})^2+p_{y}^2]}\left(                 
  \begin{array}{ccc}   
    1\\  
    0\\  
    0\\  
  \end{array}
\right)  ,              
\end{align}
where $\delta$ and $k_{0}$ are the width and the average momentum of the wave packet. The mean value of position is $\langle \vec{r}(t)\rangle=\langle g|\vec{r}(t)|g\rangle$.

\begin {figure}[H]
\centering
\includegraphics[ scale=0.70 ]{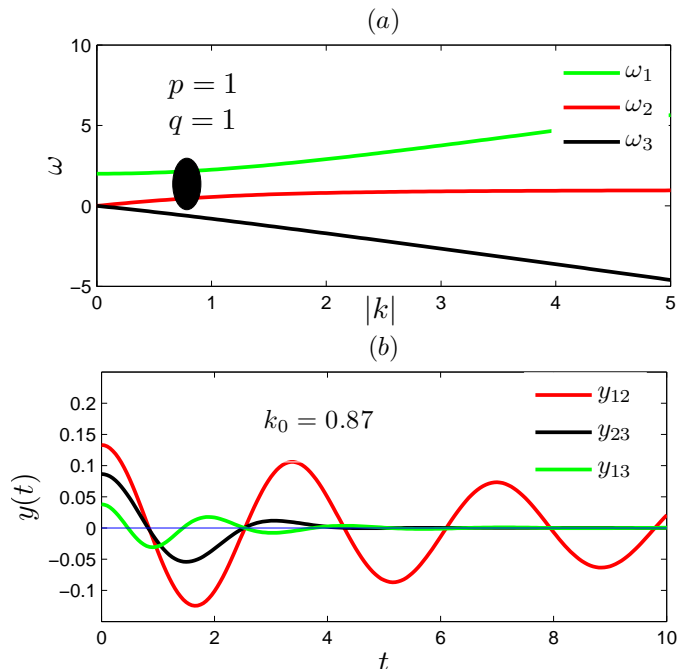}\\[0pt]
\caption{(color online). (a) The energy spectrum under both the linear and quadratic Zeeman field (p=1, q=1). The green, red and black lines denote the three energy branches $\omega_1$, $\omega_2$ and $\omega_3$, respectively. The black elliptic area is the region where the relative group velocity $v_{12}$ approaches zero. (b) The Zitterbewegung oscillations for the momentum $k_0=0.87$. The red, black and green lines denote the Zitterbewegung oscillations with the frequency corresponding to $\omega_{12}$, $\omega_{23}$ and $\omega_{13}$, respectively.    }
\label{sw2d}
\end{figure}

From Eq. (4), we can find that the kinetic energy of Hamiltonian does not contribute to the Zitterbewegung oscillation. Therefore, during the discussion of the Zitterbewegung in this section, the kinetic energy part is neglected. In this section, we take the reduced Plank constant $\hbar$, the spin-orbit coupling strength $\gamma$ and the wave packet width $\delta$ as independent fundamental units. The other derived physical quantities, such as time, momentum, energy are measured by ${\delta}/{\gamma}$, ${\hbar}/{\delta}$ and ${\gamma \hbar}/{\delta}$, respectively.

 We explore the following four cases according to the applied Zeeman field.

Case 1: The Zitterbewegung under zero Zeeman field ($p=0$ and  $q=0$).

Considering the initial wave packet (5), only the matrix element of $(y(t))_{1,1}$ has contribution to the the oscillation part of $y$ component of position operator. In the Heisenberg picture, it takes the form,
\begin{align}
y(t)_{1,1}=\frac{p_{x}}{p_{x}^{2}+p_{y}^{2}}cos(t\sqrt{p_{x}^{2}+p_{y}^{2} }).
\end{align}
Substituting the initial state, we calculate the mean value of the  oscillatory part of $y$
\begin{align}
&\langle y(t)\rangle\notag\\
&=\int_{-\infty}^{\infty}\int_{-\infty}^{\infty}d\vec{p}\frac{1}{\pi}\frac{p_{x}cos( t\sqrt{p_{x}^{2}+p_{y}^{2} })e^{{-((p_{x}-k_{0})^2+p_{y}^{2})}}}{p_{x}^{2}+p_{y}^{2}}\notag\\
&=2 e^{-k_{0}^{2}}Re[\int_{0}^{\infty}d\rho e^{-\rho^2+i t \rho}I_{1}(2 k_0\rho)],
\end{align}
where $\rho=\sqrt{p_{x}^{2}+p_{y}^{2}}$ and $I_{1}(x)$ are variable of integration and the modified Bessel function of first order, respectively. In order to capture qualitative behaviors of Zitterbewegung oscillation, we assume $ k_0\gg1$, and the asymptotic formula $I_{1}(x)\approx e^{x}/\sqrt{2\pi x}$ can be used \cite{Demikhovskii}. The integral $\int_{0}^{\infty}d\rho e^{(-\rho^2+I t \rho+2 k_0\rho)}/\sqrt{2 k_0\rho}$ can be evaluated using the method of steepest descents
\begin{align}
\langle y(t)\rangle\approx\frac{e^{-\frac{t^{2}}{4}}}{(k_{0}^{4}+\frac{k_{0}^{2}t^{2}}{4})^\frac{1}{4}}cos( k_{0}t-\frac{\theta}{2}),
\end{align}
where $\theta=\texttt{arctan}{(\frac{t}{2 k_{0}})}$ denotes a phase shift of Zitterbewegung oscillation.

 From Eq. (9), it is shown that in the case of $ k_0\gg1$, the average position in the the direction perpendicular to the average momentum $k_0$
undergos a damping oscillation with a single frequency. There are two factors which result in the damping. One is the exponentially decreasing term $e^{-\frac{t^{2}}{4}}$ which originates from the increasing spatial separation between the subpackets corresponding to the higher and lower eigenenergy branches \cite{Rusin}. Because the overlap between the subpackets gets smaller and smaller with time, the amplitude of the oscillation gets smaller and smaller. We can identify the non-vanishing relative group velocity between the subpackets corresponding to different energy branches at the average momentum $ v= \partial_{p_{x}}\omega_{12}|_{k_{0}}=1$. It is anticipated that, when the relative group velocity gets smaller, the damping would be suppressed. The other one is the term $1/(k_{0}^{4}+\frac{k_{0}^{2}t^{2}}{4})^\frac{1}{4}$, which results in a much slowly damping compared with the exponentially decreasing term. In the case of $ k_0\gg1$, the exponentially decreasing term dominate the whole damping trend before the disappearing of the Zitterbewegung. Although there exists a phase shift $\theta$ during the damping process of  Zitterbewegung oscillation, we find that, in the case of $ k_0\gg1$, the effect of the phase shift $\theta$ is not obvious before the disappearing of Zitterbewegung.

 For the limit of  $k_0\ll1$, the modified Bessel function can be expanded as taylor series $I_{1}(x)=\frac{x}{2}+\frac{x^3}{16}+\frac{x^5}{384}+O(x^7)$. By keeping only the first linear term , we get
\begin{align}
\langle y(t)\rangle&\approx2 k_0 e^{-k_{0}^{2}}\int_{0}^{\infty}d\rho e^{-\rho^2}\rho cos(t \rho)\notag\\
&= k_0 e^{-k_{0}^{2}}(1-tD(\frac{t}{2})),
\end{align}
where $D(x)=\frac{1}{2}\int_{0}^{\infty}e^{-t^2/4}\texttt{sin}(xt)dt$ is the Dawson function.

We can see from Eq. (10) that there will be no integrate Zitterbewegung oscillation when the average momentum is very small (see also panel $(b)$ in Fig. 1). We notice that a similar result is found in $4\times 4$ Luttiger Hamiltonian by Demikhovskii \textsl{et al}. \cite{Demikhovskii2}.

From the above approximate formulas, the amplitude of Zitterbewegung oscillation $y(t=0)$ can be obtained for the two limit case. In the case of  the average momentum $k_0\ll1$, it is proportional to the average momentum. When $k_0\gg1$, the Zitterbewegung amplitude is inversely proportional to the average  momentum. We numerically investigate the amplitude as a function of the average momentum in the panel $(a)$ of Fig.1. In panel $(a)$, there exists a maximum value of the amplitude on the curve. The maximum value occur at around $k_0\sim1$.

Case 2: The Zitterbewegung under the linear Zeeman field ($p\neq0$ and  $q=0$).

The oscillatory part of $y(t)$ in the Heisenberg picture is
\begin{align}
(y(t))_{1,1}=\frac{p_{x}}{p_{x}^{2}+p_{y}^{2}+p^2}cos( t\sqrt{p_{x}^{2}+p_{y}^{2}+p^2}).
\end{align}

 In principle, one could be able to get a asymptotic result though a similar calculations as the case of zero Zeeman field. However the resulting expression is so long and cumbersome that we could not get clear physical meaning from it. Inspired by the case of zero Zeeman field, we will give approximate expression to fit the  data obtained from the exact numerical integral under conditions of $ k_0\gg1$. The approximate formula is

\begin{align}
&\langle y(t)\rangle=\langle g|y(t)|g\rangle\notag\\
&=2e^{-k_{0}^{2}}Re[\int_{0}^{\infty}d\rho \frac{\rho^{2}e^{-\rho^2+I t \sqrt{\rho^2+p^2}}}{\rho^2+p^2}I_{1}(2 k_0\rho)]\notag\\
&\approx\frac{k_{0}^2}{k_{0}^2+p^2}\frac{ e^{-\frac{v_{12}^{2}t^{2}}{4}}}{(k_{0}^{4}+\frac{k_{0}^{2}v_{12}^{2}t^{2}}{4})^\frac{1}{4}}\texttt{cos}(\omega_{12}t),
\end{align}
where $v_{12}=\partial_{p_x}\omega_{12}|_{k_{0}}=\frac{k_{0}}{\sqrt{k_{0}^{2}+p^2}}$ is the relative group velocity at the average momentum and $\omega_{12}=\sqrt{k_{0}^{2}+p^2}$ is the energy difference between different energy branches at the average momentum.
We can see from Eq. (12) that the amplitude is suppressed by the applied linear Zeeman field. With the increase of the linear Zeeman field, the decaying trend is suppressed. The reason is that with the increase of linear Zeeman field, the relative velocity between sub-packets corresponding to different energy branches at the average momentum $ v_{12}= \frac{k_0}{\sqrt{ k_{0}^{2}+p^2}} $ gets smaller. Then, the sustained coherence between sub-packets lead to the suppression of decaying.
For some specific parameters, we depict the Zitterbewegung oscillations with only the linear Zeeman field in Fig. 2.

Case 3: The Zitterbewegung under the quadratic Zeeman field ($p=0$ and  $q\neq0$).

The oscillatory part $y(t)_{1,1}$of the position operator in the Heisenberg picture  is
\begin{align}
y(t)_{1,1}=\frac{p_{x}
   \omega _{23} \cos (t \omega _{12})}{2 \omega
   _{13} (p_{x}^{2}+p_{y}^2)}+\frac{p_{x}
   \omega _{12} \cos (t \omega _{23})}{2 \omega
   _{13} (p_{x}^{2}+p_{y}^2)},
\end{align}
where $\omega _{23}=\frac{1}{2}(q+\sqrt{4(p_{x}^{2}+p_{y}^{2})+q^2})$,  $\omega _{12}=\frac{1}{2}(-q+\sqrt{4(p_{x}^{2}+p_{y}^{2})+q^2})$, $\omega _{13}=\sqrt{4(p_{x}^{2}+p_{y}^{2})+q^2}$.

 By using the similar approximation, the average position along y direction is approximated by

\begin{align}
\langle y(t)\rangle&=\langle g|y(t)|g\rangle\notag\\
&=y_{12}+y_{23}\notag\\
&\approx\frac{\omega_{2,3}(k_0)}{2\omega_{13}}\frac{ e^{-\frac{v_{12}^{2}t^{2}}{4}}}{(k_{0}^{4}+\frac{k_{0}^{2}v_{12}^{2}t^{2}}{4})^\frac{1}{4}}\texttt{cos}(\omega_{12}t)\notag\\
&+\frac{\omega_{12}(k_0)}{2\omega_{13}}\frac{ e^{-\frac{v_{23}^{2}t^{2}}{4}}}{(k_{0}^{4}+\frac{k_{0}^{2}v_{23}^{2}t^{2}}{4})^\frac{1}{4}}\texttt{cos}(\omega_{23}t
),
\end{align}
where $y_{ij}$ is the Zitterbewegung oscillation with frequency which is  energy difference $\omega_{ij}$ between the energy branches $i$ and $j$ at the average momentum. $v_{12}=\partial_{p_x}\omega_{12}|_{k=k_{0}}=\frac{k_{0}}{\sqrt{k_{0}^{2}+q^2}}$,  $v_{23}=\partial_{p_x}\omega_{23}|_{k=k_{0}}=\frac{k_{0}}{\sqrt{k_{0}^{2}+q^2}}$ are relative group velocities between different energy branches at the central momentum.

In the presence of only the quadratic Zeeman filed, the Zitterbewegung oscillation split into two oscillations with different amplitudes and frequencies (see also Fig. 3). The amplitude of the Zitterbewegung oscillation with higher frequency is smaller than the other one. Because of the equal relative group velocities at the central momentum $v_{12}=v_{23}$, the two Zitterbewegung oscillations have the same exponential decreasing factor and the decaying trends for the two Zitterbewegung oscillations are the same on the whole. With the increase of the quadratic Zeeman field, the splitting of both the amplitudes and frequencies between the two Zitterbewegung  oscillations gets more evident.

Case 4: The Zitterbewegung under both the linear and quadratic Zeeman field ($p\neq0$ and  $q\neq0$).

In the presence of both the linear and quadratic zeeman field, the analytical expression of the position operator in the  Heisenberg picture is quite long and cumbersome. We depict the Zitterbewegung oscillations with some specific parameters, e.g. $p=1$, $q=1$ and $k_0=0.87$ in Fig.4. We can see that there are usually three frequencies when both the linear and quadratic Zeeman field exist.
We find that when the relative group velocity $v_{ij}= \partial_{p_{x}}\omega_{ij}|_{k_0}$ between different eigen-states at the central momentum approaches zero, the mean position will damp much slowly. As shown in the bottom panel of Fig.4, when the central momentum is chosen near the elliptic region and the relative velocity between sub-packets nearly vanish, the decaying trend of the Zitterbewegung oscillation corresponding to $\omega_{12}$ is suppressed greatly compared with the two others. The slowly decaying amplitude may be favorable to the observation of Zitterbewegung in experiments.

\section{The Zitterbewegung effect in harmonic trap }
As shown in the above section, the amplitude of Zitterbewegung  oscillation in free space usually decays with time. In the trapped system, the oscillatory behaviors have unique characteristics compared with that in free space. Firstly, due to the confinement of trap, the atoms could not escape from the trap. Hence, the subpackets corresponding to different eigenenergies would always keep coherent, resulting in non-decaying Zitterbewegung oscillations. Secondly, due to the existence of infinite energy levels in the trap, the Zitterbewegung oscillation shows very complicated features.

In this section, we adopt the natural units for harmonic oscillators. The length, mass and time are measured by $\sqrt{\hbar/m\omega}$, $m$ and $1/\omega$ respectively. The other physical quantities, such as, energy, momentum, velocity are measured by $\hbar\omega$, $\sqrt{\hbar m \omega}$ and $\sqrt{\hbar \omega/m}$, respectively.

Similar to that in Ref. \cite{David}, we decompose the time evolution operator as $ U=e^{-iHt}=\sum_{l}e^{-iE_{l}t}|l\rangle\langle l|$, where $|l\rangle$ is the eigenstates of Hamiltonian and $E_{l}$ the corresponding eigenenergy. We generalize Eq. (3) to the trapped system and obtain the position operator in Heisenberg picture as
\begin{align}
& \vec{r}(t)=\vec{r}(0)+\sum_{k}Z_{k;k}+\sum_{k,l\neq k}e^{i\omega_{kl}t}\vec{Z}_{k;l},
\end{align}
where $\vec{Z}_{k;l}=|k\rangle\langle k|\vec{r}|l\rangle\langle l|$ is the so-called Zitterbewegung amplitude operator, and $\omega_{kl}$ is energy difference between eigenenergies.
To calculate the average value of position operator, the eigenstates and eigenenergies are required. Before the numerical calculation of the eigen-equations, we will discuss the symmetries of Hamiltonian. The Hamiltonian have rotational symmetries along $z$ axial direction. Thus the total angular momenta along $z$ direction $J_{z}=L_{z}+F_{z}$ is a good quantum number. We will label the eigenstates and energies with good quantum number $j_z=m$. In the polar coordinates $(\rho, \theta)$, the eigen-function can be written in the following form

\begin{align}
|\psi_{m}(\rho,\theta)\rangle=\left(                 
 \begin{array}{ccc}   
   \phi_{1}(\rho)\frac{e^{i(m-1)\theta}}{\sqrt{2\pi}}\\  
  \phi_{0}(\rho)\frac{e^{im\theta}}{\sqrt{2\pi}}  \\  
   \phi_{-1}(\rho)\frac{e^{i(m+1)\theta}}{\sqrt{2\pi}}    \\
\end{array}\right),
\end{align}
with $j_{z}=m$.

In addition to the rotational symmetry, there is time reversal symmetry in the absence of the Zeeman filed. The time reversal operator is expressed as $T=UK$ with $U=exp(-i\pi F_{y})$ and $K$ the complex conjugate operation.  Its matrix form is
\begin{align}
T=\left(                 
 \begin{array}{ccc}   
   0   &    0    &  1 \\  
   0 &    -1    &    0    \\  
   1     &  0   &   0
\end{array}\right)K.
\end{align}
Then the time reversal state can be obtained as $|\psi_{-m}(\rho,\theta)\rangle=T|\psi_{m}(\rho,\theta)\rangle$ with degenerate eigenenergies $E_{-m,l}=E_{m,l}$. The eigenstates are doubly degenerate except for the states corresponding to $m=0$.

\begin{figure}[tbp]
\begin{center}
\includegraphics[ scale=0.65 ]{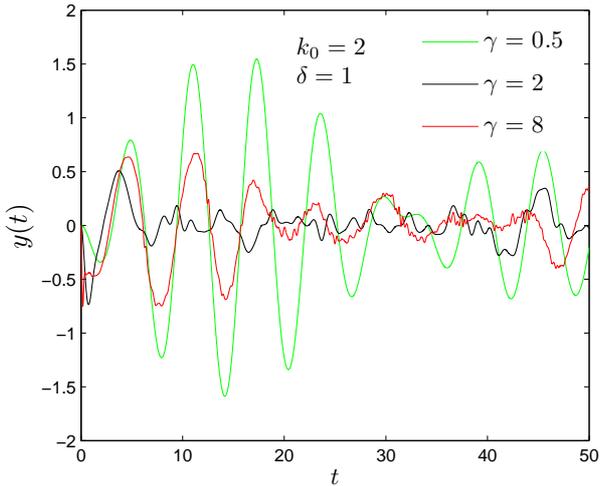}\\[0pt]
\end{center}
\caption{(color online). The Zitterbewegung oscillations with a Gaussian wave packet as an initial state in the trap for the wave packet width $\delta=1$, the momentum $k_0=2$. The green, black and red lines correspond to the spin-orbit coupling strength $\gamma=$ 0.5, 2 and 8, respectively.  }
\label{sw2d}
\end{figure}

We will work with the basis of 2D harmonic oscillator, and the basis can also be introduced by two independent operators in terms of operators of $a_{x(y)}$ \cite{Cohen-Tannoudji}
\begin{align}
a_{d}=&\frac{1}{\sqrt{2}}(a_{x}-ia_{y}),\notag \\
a_{g}=&\frac{1}{\sqrt{2}}(a_{x}+ia_{y}),
\end{align}
where $a_{x(y)}$ is annihilation operators  of the $x(y)$ direction. The position and momentum operators can be expressed in terms of the operators in Eq.(18) and their corresponding adjoint operators. For example, the $y$ component of the position is $y=\frac{i}{2}(a_{d}-a_{d}^{+}-a_{g}+a_{g}^{+})$. The harmonic oscillator basis can be written as
\begin{align}
|\chi_{n_{d},n_{g}}\rangle=\frac{1}{\sqrt{(n_{d})!(n_{g})!}}(a_{d}^{+})^{n_{d}}(a_{g}^{+})^{n_{g}}|0,0\rangle,
\end{align}
where $|0,0\rangle$ is the ground state of a 2D harmonic oscillator. The 2D harmonic oscillator basis can be expressed in the coordinate space as
\begin{align}
\phi_{n,m}(\rho,\theta) =\langle \vec{r}|\chi_{n_{d},n_{g}}\rangle=R_{n,m}(\rho)\frac{e^{im\theta}}{\sqrt{2\pi}},
\end{align}
where  $n=n_{g}$, $m=n_{d}-n_{g}$, $R_{n,m}(\rho)=(-1)^{n}\sqrt{\frac{2(n!)}{(n+|m|)!}}   \rho^{|m|}e^{-\frac{\rho^2}{2}}L_{n}^{|m|}(\rho^2)$, $L_{n}^{|m|}(x)=\Sigma_{k=0}^{n}C_{n+|m|}^{n-k}\frac{(-x)^{k}}{k!}$ is the associated Laguerre polynomial and $C_{n}^{k}=\frac{n!}{(n-k)!k!}$ is  the binomial coefficient.

From the expression of the position operators in terms of $a_{g(d)}$ and their adjoint operators, there exist an important selection rule for $Zitterbewegung$ $amplitude$ operators
\begin{align}
\vec{Z}_{m,l;m',l'}=|m,l\rangle\langle m,l|\vec{r}|m',l'\rangle\langle m',l'|=0, \notag\\
(m-m'\neq\pm1),
\end{align}
where $l$ labels the eigenstates belonging to the same quantum $m$.
It is shown that only for $m=m\prime\pm1$ , there exist no-vanishing $Zitterbewegung$ $amplitude$. From the above equation, it is found that the second term in Eq. (15) vanishes, i.e.
$\vec{Z}_{k;k}=\vec{Z}_{m,l;m,l}=0$.

One can express the Hamiltonian in the form of matrix by adopting the harmonic oscillator basis. The corresponding eigenenergies and eigenstates can be  obtained through direct numerical diagonalization \cite{Ramachandhran}.

Now we investigate the Zitterbewegung oscillatory characteristics in the trap. First of all, as shown by equation (15), when the the initial state is the superposition of two eigenstates, there would be a an oscillation with one single frequency which is the difference of two eigen-energies. The amplitude of the oscillation does not decay. This is because that the components corresponding to two eigenenergies in the initial state always keep coherent in the trap as stated before.

Next, we calculated the Zitterbewegung oscillations when an Gaussian wave packet with non-vanishing average momentum is used as an initial state. The projections of the initial state onto the harmonic basis are required in the calculation the average value of the position.  The detailed calculations of the projection are listed in the appendix. From Eq. (15), we know that the number of the frequencies of the oscillation involved in Zitterbewegung oscillation is usually arbitrary when the initial state is chose in a completely arbitrary fashion. Using the Gaussian wave packet as initial state, we find that there are indeed a lot of frequencies in the Zitterbewegung oscillation.

For some specific parameters, the Zitterbewegung oscillations in the trap are shown in Fig. 5. From Fig. 5, we can see that, for both the weak and strong spin-orbit coupling, the Zitterbewegung oscillations display the harmonic oscillator characteristics within short time interval (20 time unit in Fig.5). This is because when the spin-orbit coupling approaches to zero, the energy-spectrum approach the harmonic oscillator spectrum. In the case of the strong spin-orbit limit,  the eigenenergies can be approximated by $E_{m,n}=[-\gamma^{2}+2n+1+m^2/\gamma^{2}]/2$, which is analogous to the energy spectrum formula of two component atoms with strong spin-orbit coupling and harmonic trap \cite{WU,Hu}. Compared with the two components atoms, the ground state is not degenerate. The energy spectrum in strong spin-orbit coupling limit form landau level-like spectrum. Thus, in cases of both weak and strong spin-orbit coupling limit, the energy level spacing approaches $\hbar \omega$.
 With the increase of the time, the out of phase oscillations begin to appear. Meanwhile, the oscillations with high frequency begin to manifest themselves (see the red line in Fig. 5).

In the presence of the intermediate-strength spin-orbit coupling, the Zitterbewegung shows quite complicatedly and irregularly oscillatory characteristics. It is due to the complicated energy spectrum and coexistence of oscillations with various frequencies in Zitterbewegung.

\begin{figure}[tbp]
\begin{center}
\includegraphics[ scale=0.70 ]{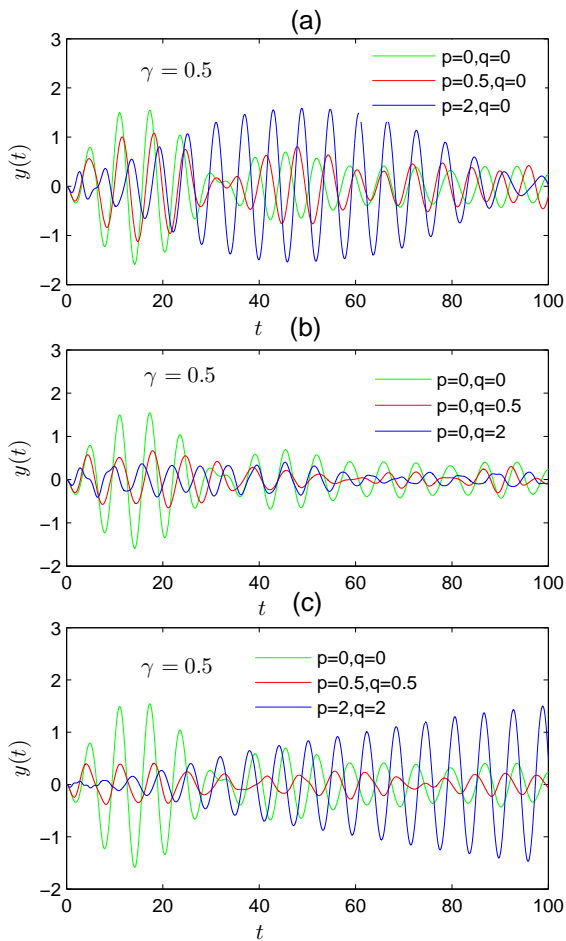}\\[0pt]
\end{center}
\caption{(color online). The Zitterbewegung oscillations in the presence of both the harmonic trap and  the Zeeman field with the wave packet width $\delta=1$, the average momentum $k_0=2$ and the spin-orbit coupling strength $\gamma=0.5$, respectively. The green, red and blue lines correspond to the  Zitterbewegung oscillations for various Zeeman parameters.  }
\label{sw2d}
\end{figure}

\begin{figure}[tbp]
\begin{center}
\includegraphics[ scale=0.70 ]{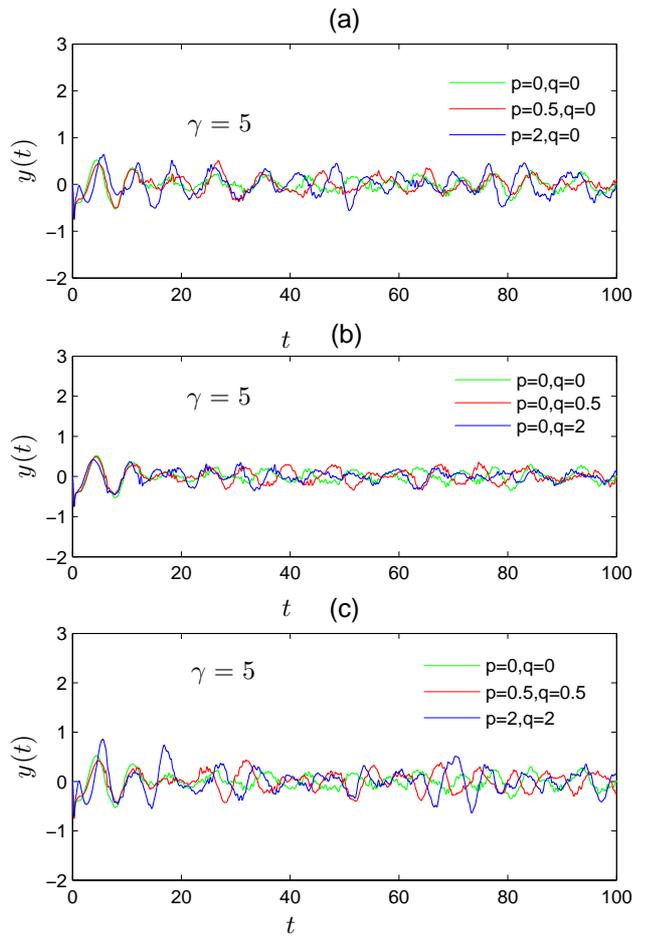}\\[0pt]
\end{center}
\caption{(color online). The Zitterbewegung oscillations in the presence of both the harmonic trap and the Zeeman field with the wave packet width $\delta=1$, the average momentum $k_0=2$ and the spin-orbit coupling strength $\gamma=5$, respectively.   }
\label{sw2d}
\end{figure}

%
We explore the Zitterbewegune oscillation in the presence of both the Zeeman field and the harmonic trap. For the weak spin-orbit coupling, the top panel in Fig. 6 shows that the Zitterbewegung effect is manifested as  beat oscillation. The beat period gets larger and larger with the increase of the linear Zeeman filed. The quadratic Zeeman field usually suppress the amplitude of the Zitterbewegung oscillation, as shown by the the middle panel in Fig. 6. In the presence of both the linear and quadratic Zeeman fields, we can see that the beat period gets even much larger compared with the case where only the linear Zeeman field is included as shown the bottom panel in Fig. 6 (see the blue lines). For strong spin-orbit coupling, the amplitude of the Zitterbewegune oscillation is usually smaller than that of weak spin-orbit coupling as shown in Fig. 7. At the same time the high oscillations is manifested more evidently.

\begin{figure}[tbp]
\begin{center}
\includegraphics[ scale=0.70 ]{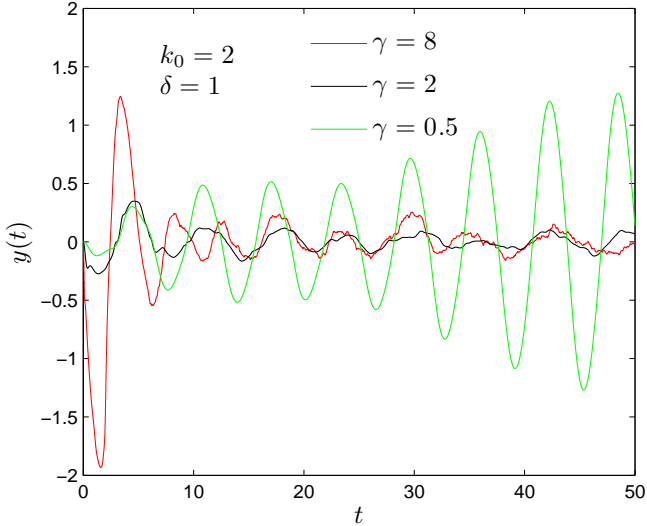}\\[0pt]
\end{center}
\caption{(color online). The Zitterbewegung oscillations in the presence of the both harmonic trap and the interaction between atoms with the wave packet width $\delta=1$, the momentum $k_0=2$, respectively.   }
\label{sw2d}
\end{figure}

\section{Effect of interaction on the Zitterbewegung oscillation}
 In this section, we numerically simulate the dynamical evolution of the spin-orbit coupled $^{23}$Na atoms by solving the corresponding GP equation. Taking the axial trap frequency $\omega_z=2\pi\times800$Hz, the transverse trap frequency $\omega_\bot=2\pi\times90$Hz, the total number of atoms N$=10^4$ and the experimental values of scattering length of $^{23}$Na atoms $a_0=46a_B$, $a_2=52a_B$ ($a_B$ is the Bohr radius), the corresponding two dimensionless interaction parameters $c_{0}=178.90$, $c_{2}=7.16$ (in natural units of harmonic oscillator). The effect of interaction on the Zitterbewegung oscillation in the trap is considered in Fig. 8. Comparing it with Fig.5 (without interaction), the beat oscillation period get longer for weak spin-orbit coupling. For intermediate and strong spin-orbit coupling strength, the oscillatory amplitude for a long time is slightly suppressed.  Even though the ZB effect originates from the single particle Hamiltonian, it is still robust in the presence of interaction between atoms. Therefore, the ZB oscillation is an universal phenomenon in the dynamical evolution of the spin-orbit coupled atoms.

 The Zitterbewegune oscillation is accompanied by a damped oscillation of the probability density  with time for each component \cite{Rusin}. The oscillatory properties of position operator, such as the decaying trend, the oscillatory frequency, e.g, can be reflected by the damped oscillation of the number of atoms.  The oscillation of the number can be obtained through time-of-flight measurement \cite{Wang2}. The parameters involved in the Hamiltonian of atoms can be precisely controlled in cold atoms experiments. Taking $^{23}$Na atoms for example, one can adopt the typical experiment parameters, such as the wave length of Raman laser $\lambda\sim600nm$, which corresponds to a spin-orbit strength $\gamma=\sqrt{2}\pi\hbar/(m\lambda)\approx1.93 cm/s$ ($m=3.82\times10^{-26}kg$ for $^{23}$Na).
One can prepare a gaussian packet of cold atoms and apply a series of time-of-flight measurement for various values of time. Then it is expected to obtain the decaying oscillation of the number of atoms.


\section{Summary}
In summary, we have investigated the Zitterbewegung effect in spin-$1$ atoms in the presence of the Zeeman field and the external harmonic trap. It is shown that the Zitterbewegung oscillations could be greatly affected by the external Zeeman field and the trap. The external Zeeman field could suppress or enhance the Zitterbewegung amplitude and change the frequencies of oscillation. In addition, multi-frequency oscillations could appear in the presence of the Zeeman field. Through adjusting both the linear and quadratic Zeeman field properly, we could obtain a much slowly damping  Zitterbewegung oscillation. In the presence of harmonic trap, there would be the Zitterbewegung oscillation without damping and with arbitrary number of frequencies. The Zitterbewegung in the trap displays very complicated and irregular oscillation characteristics due to its complicated energy spectrum. It is shown that the Zitterbewegung oscillation is robust even in the presence of interaction between atoms.

The present work should help us understand the roles of external Zeeman field and trap in the Zitterbewegung oscillations. The investigation of the effects induced by the external field and trap on Zitterbewegung oscillation opens up new possibilities for the manipulation and control of the Zitterbewegung oscillation. We anticipate the Zitterbewegung effect be detected in the widely-studied cold atoms experiments with spin-orbit coupling interactions.

\noindent{\bf Acknowledgements:}
This work was supported by the NKBRSFC under grants Nos. 2011CB921502, 2012CB821305, 2009CB930701, 2010CB922904, and NSFC under grants Nos. 10934010 and NSFC-RGC under grants Nos. 11061160490 and 1386-N-HKU748/10.

\appendix

\section{The projections of the Gaussian packet onto the harmonic oscillator basis }
In this appendix, we give the calculation of the projections of the initial Gaussian wave packet onto the harmonic oscillator basis, which is required in the calculation of the average values of position operator.
The initial wave packet reads:

\begin{align}
|g\rangle=\frac{1}{\sqrt{\pi\delta^{2}}}e^{-\frac{(x^{2}+y^{2})}{2\delta^{2}}}e^{ik_0x},
\end{align}
where $\delta$ denotes the width of the wave packet and $k_{0}$ is the average momentum of wave packet.

The projection onto the harmonic oscillator basis is
\begin{align}
\langle\phi_{n,m}|g\rangle=\frac{2(-1)^{n}i^{|m|}}{\delta}\sqrt{\frac{n!}{(n+|m|)!}}\times\notag\\
\int_{0}^{+\infty}\rho^{|m|+1}e^{-\sigma^{2}\rho^{2}}L_{n}^{|m|}(\rho^2)J_{|m|}(k_{0}\rho)d\rho,
\end{align}

where $|\phi_{n,m}\rangle=R_{n,m}(\rho)\frac{e^{im\theta}}{\sqrt{2\pi}}$, $\sigma^2=\frac{1}{2}(1+\frac{1}{\delta^2})$ and $J_{|m|}(x)$ is the Bessel functions of order $|m|$.

The associated Laguerre polynomial consists of  various monomials. The integrals corresponding to the monomials can be obtained through an integral identity
\begin{align}
&\int_{0}^{+\infty}\rho^{\mu-1}e^{-p^{2}\rho^{2}}J_{\nu}(a\rho)d\rho\notag\\
&=\frac{\Gamma(\frac{\mu+\nu}{2})}{2p^{\mu}\Gamma(\nu+1)}(\frac{a}{2p})^{\nu}e^{-\frac{a^{2}}{4p^{2}}}{_1}F_{1}(\frac{\nu-\mu}{2}+1;\nu+1;\frac{a^{2}}{4p^{2}}), \end{align}
where $\Gamma(z)$ is Euler's Gamma function and ${_1}F_{1}(\alpha;\beta;x)$ is the generalized hypergeometric functions \cite{Wang1}.
Combining Eq. (A2) with Eq. (A3), we can obtain the projections by summing up all integral values corresponding to the monomials.
The Zitterbewegung amplitude operators $\vec{Z}_{k;l}$ in Eq. (15) are also calculated within the harmonic oscillator basis and the mean value of the position operator can be obtained correspondingly.


\begin{references}















%

\bibitem{Schrodinger} E. Schr\"{o}dinger, Sitzungsber. Preuss. Akad. Wiss. Phys.
Math. Kl. 24, 418 (1930)
%
\bibitem{Rusin2} T. M. Rusin, and W. Zawadzki, Phys. Rev. B \textbf{80}, 045416 (2009).
\bibitem{Schliemann1} J. Schliemann, D. Loss, and R. M. Westervelt,  Phys. Rev. Lett. \textbf{94}, 206801 (2005).
\bibitem{Jozsef} J. Cserti and G. D\'{a}vid, Phys. Rev. B \textbf{74}, 172305 (2006).
\bibitem{Zhang2} X. D. Zhang,  Phys. Rev. Lett. \textbf{100}, 113903 (2008).
\bibitem{Lamata} L. Lamata, J. Le\'{o}n, T. Sch\"{a}tz, and E. Solano,  Phys. Rev. Lett. \textbf{98}, 253005 (2007).


\bibitem{Gerritsma} R. Gerritsma, G. Kirchmair, F. Z\"{a}hringer, E. Solano, R. Blatt and C. F. Roos, Nature 463, 68 (2010).
\bibitem{Schliemann} J. Schliemann,  New J. Phys. \textbf{10}, 043924 (2008).









\bibitem{Jiang} Z. F. Jiang, R. D. Li, S. C. Zhang, and W. M. Liu, Phys. Rev. B \textbf{72}, 045201 (2005).
\bibitem{Shen} S. Q. Shen,  Phys. Rev. Lett. \textbf{95}, 187203 (2005).

\bibitem{Winkler} R. Winkler, U. Z\"{u}licke  and J. Bolte, Phys. Rev. B \textbf{75}, 205314 (2007).
\bibitem{Lee} M. Lee and C. Bruder, Phys. Rev. B \textbf{72}, 045353 (2005).

\bibitem{Wang} Z. Y. Wang and C. D. Xiong, Phys. Rev. A \textbf{77}, 045402 (2008).



\bibitem{Lin1} Y. J. Lin, K. J. Garc\'{\i}a and I. B. Spielman, Nature \textbf{471}, 83 (2011).
\bibitem{Cheuk} L. W. Cheuk, A. T. Sommer, Z. Hadzibabic, T. Yefsah, W. S. Bakr, and M. W. Zwierlein,  Phys. Rev. Lett. \textbf{109}, 095302 (2012).
\bibitem{Lin} Y. J. Lin, R. L. Compton, A. R. Perry, W. D. Phillips, J. V. Porto, and I. B. Spielman,  Phys. Rev. Lett. \textbf{102},  130401 (2009).
















\bibitem{Wang2} P. J. Wang, Z. Q. Yu, Z. K. Fu, J. Miao, L. H. Huang, S. J. Chai, H. Zhai and J. Zhang,  Phys. Rev. Lett. \textbf{109},  095301 (2012).

\bibitem{Juzeliunas} G. Juzeli\={u}nas, J. Ruseckas and J. Dalibard, Phys. Rev. A \textbf{81}, 053403 (2010).



\bibitem{Zhai} C. Wang, C. Gao, C. M. Jian, and H. Zhai,  Phys. Rev. Lett. \textbf{105},  160403 (2010).

\bibitem{Ho} T. L. Ho and S. Zhang,  Phys. Rev. Lett. \textbf{107},  150403 (2011).

\bibitem{Yip} S.-K. Yip, arXiv: 1008. 2263  (2011).

\bibitem{Hu} H. Hu, B. Ramachandhran, H. Pu, and X. J. Liu,  Phys. Rev. Lett. \textbf{108},  010402 (2012).

\bibitem{Ramachandhran} B. Ramachandhran, B. Opanchuk, X. J. Liu, H. Pu, P. D. Drummond, and H. Hu, Phys. Rev. A \textbf{85}, 023606 (2012).
\bibitem{Santos} S. Sinha, R. Nath, and L. Santos,  Phys. Rev. Lett. \textbf{107},  270401 (2011).
\bibitem{Xu} X. Q. Xu and J. H. Han,  Phys. Rev. Lett. \textbf{107},  200401 (2011).
\bibitem{Zhou} X. F. Zhou, J. Zhou, and C. J. Wu, Phys. Rev. A \textbf{84}.  063624  (2011).
\bibitem{WU}  C. J. Wu, I. M. Shem, X. F. Zhou,  Chin. Phys. Lett. \textbf{28},  097102 (2011).
\bibitem{Guo} S.-W. Su, I.-K. Liu, Y.-C. Tsai, W. M. Liu, and S.-C. Gou, Phys. Rev. A \textbf{86}.  023601  (2012).





\bibitem{Vaishnav} J. Y. Vaishnav and C. W. Clark,  Phys. Rev. Lett. \textbf{100}, 153002 (2008).

\bibitem{Zhang} Q. Zhang, J. Gong and C. H. Oh, Phys. Rev. A \textbf{81}.  023608  (2010).




\bibitem{Zhang1}  D. W. Zhang, Z. D. Wang, S. L. Zhu, arXiv: 1203.5949v1 (2012).
\bibitem{Merkl} M. Merkl, F. E. Zimmer, G. Juzeli\={u}nas and P. \"{o}hberg, Europhys. Lett.  \textbf{ 83}, 54002 (2008).
\bibitem{David} G. D\'{a}vid and J. Cserti, Phys. Rev. B \textbf{81}, 121417 (2010).






\bibitem{Demikhovskii} V. Y. Demikhovskii, G. M. Maksimova, and E. V. Frolova, Phys. Rev. B \textbf{78}, 115401 (2008).

\bibitem{Rusin} T. M. Rusin, and W. Zawadzki, Phys. Rev. B \textbf{76}, 195439 (2007).

\bibitem{Demikhovskii2} V. Y. Demikhovskii, G. M. Maksimova, and E. V. Frolova, Phys. Rev. B \textbf{81}, 115206 (2010).



\bibitem{Cohen-Tannoudji} C. C. Tannoudji, B. Diu, F. Lalo\"{e}, Quantum Mechanics, Vol. I (Hermann and John Wiley \& Sons, 1977).

























\bibitem{Wang1} Z. X. Wang and D. R. Guo, Special Functions, (World Scientific, Singapore, 1989).














%
%
%
%
%
%
%
%
%
%
%
%
%
%
%
%
%
%
%
%
%
%
%
%
%
%
%
%
%
\end{references}
\end{document}